\pdfoutput=1
\documentclass[aps,prd,twocolumn,superscriptaddress,preprintnumbers,floatfix,nofootinbib]{revtex4-2}

\usepackage{amsmath}
\usepackage{amsfonts}
\usepackage{amssymb}	
\usepackage{graphicx}
\usepackage{enumitem}
\usepackage[dvipsnames]{xcolor}
\usepackage[colorlinks,urlcolor=NavyBlue,citecolor=NavyBlue,linkcolor=NavyBlue,pdfusetitle]{hyperref}
\usepackage[all]{hypcap}
\usepackage[T1]{fontenc}
\usepackage[utf8]{inputenc}
\usepackage{microtype}
\usepackage{mathrsfs}
\usepackage{siunitx}
\usepackage[normalem]{ulem} %

\graphicspath{{../}{../figures}}

\newcommand{\scri}{\mathscr{I}}
\newcommand{\modelName}{\texttt{NRSur3dq8\_RD}}
\newcommand{\rmd}{\mathrm{d}}

\usepackage{orcidlink}
\newcommand{\OleMiss}{\affiliation{Department of Physics and Astronomy,
		University of Mississippi, University, Mississippi 38677, USA}}

\newcommand{\UMassD}{\affiliation{Department of Mathematics and Center 
    for Scientific Computing \& Visualization Research, University of 
    Massachusetts, Dartmouth, MA 02747}}
\newcommand{\URI}{\affiliation{Department of Physics and Center for 
    Computational Research, East Hall, University of Rhode Island, Kingston, 
    RI 02881}}
\newcommand{\CornellLepp}{\affiliation{Laboratory for Elementary Particle Physics,
		Cornell University, Ithaca, New York 14853, USA}}
\newcommand{\CornellPhysics}{\affiliation{Department of Physics,
		Cornell University, Ithaca, New York 14853, USA}}
\newcommand{\Cornell}{\affiliation{Cornell Center for Astrophysics and Planetary Science,
		Cornell University, Ithaca, New York 14853, USA}}
\newcommand{\Caltech}{\affiliation{Theoretical Astrophysics 350-17,
		California Institute of Technology, Pasadena, California 91125, USA}}
\newcommand{\MaxPlanck}{\affiliation{Max Planck Institute for Gravitational Physics (Albert Einstein Institute), Am M{\"u}hlenberg 1, Potsdam 14476, Germany}}

\begin{document}

\title{High-Precision Ringdown Surrogate Model for Non-Precessing Binary Black Holes}

\author{Lorena \surname{Magaña Zertuche}
	\orcidlink{0000-0003-1888-9904}}
\email{lmaganaz@go.olemiss.edu}
\OleMiss
\author{Leo C.~Stein
	\orcidlink{0000-0001-7559-9597}}
\OleMiss
\author{\\Keefe Mitman
	\orcidlink{0000-0003-0276-3856}}
\Caltech
\author{Scott E. Field
  \orcidlink{0000-0002-6037-3277}}
\UMassD
\URI
\author{Vijay Varma
  \orcidlink{0000-0002-9994-1761}}
\UMassD
\author{Michael Boyle
	\orcidlink{0000-0002-5075-5116}}
\Cornell
\author{Nils Deppe
	\orcidlink{0000-0003-4557-4115}}
\CornellLepp
\CornellPhysics
\Cornell
\author{Lawrence E.~Kidder~%
	\orcidlink{0000-0001-5392-7342}}
\Cornell
\author{Jordan Moxon
	\orcidlink{0000-0001-9891-8677}}
\Caltech
\author{Harald P.~Pfeiffer~%
	\orcidlink{0000-0001-9288-519X}}
\MaxPlanck
\author{Mark A. Scheel
	\orcidlink{0000-0001-6656-9134}}
\Caltech
\author{Kyle C.~Nelli
	\orcidlink{0000-0003-2426-8768}}
\Caltech
\author{William Throwe
	\orcidlink{0000-0001-5059-4378}}
\Cornell
\author{Nils L.~Vu
	\orcidlink{0000-0002-5767-3949}}
\Caltech

\hypersetup{pdfauthor={\surname{Magaña Zertuche} et al.}}

\date{\today}

\begin{abstract}
	Highly precise and robust waveform models are required as improvements in 
	detector sensitivity enable us to test general relativity with more precision than 
	ever before. In this work, we introduce a spin-aligned surrogate ringdown model.
	This ringdown surrogate, \modelName, is built with numerical waveforms produced 
	using Cauchy-characteristic evolution. In addition, these waveforms are in the 
	superrest frame of the remnant black hole allowing us to do a correct analysis 
	of the ringdown spectrum.
  The novel prediction of our surrogate model is complex-valued
  quasinormal mode (QNM) amplitudes, with median relative errors of
  $10^{-2}-10^{-3}$ over the parameter space. Like previous remnant
  surrogates, we also predict the remnant black hole’s mass and spin.
  The QNM mode amplitude errors translate into
  median errors on ringdown waveforms of $\sim10^{-4}$. The high accuracy and QNM mode
	content provided by our surrogate will enable high-precision ringdown analyses
	such as tests of general relativity. Our ringdown model is publicly
  available through the python package \texttt{surfinBH}.

\end{abstract}

\maketitle
\section{Introduction}
After the cataclysmic collision of two black holes, the remnant black hole
vibrates at different frequencies in what is known as the ringdown phase. These
quasinormal mode (QNM) frequencies are dictated by the mass and spin of the
final black hole. Understanding the corresponding amplitudes of these QNMs
allows us to model the aftermath of a binary black hole (BBH) 
merger~\cite{Teukolsky:1973,Detweiler:1980gk,Leaver:1985ax,Dolan:2009nk}. However, to
estimate the QNM amplitudes, we must rely on numerical relativity (NR)
waveforms. Yet producing a single numerical relativity waveform describing a specific 
BBH system is computationally expensive and can take weeks to months to run.
An additional layer of complication is added due to the large BBH parameter
space --- consisting of both BH masses and spin vectors among other parameters.
Due to this, it is challenging to densely populate the full parameter space with NR 
waveforms and, therefore, it presents a difficulty for accurate parameter inference 
with current ground-based detectors. 

Nevertheless, there are several BBH waveform models that help ameliorate this 
problem. A popular framework used in dealing with the general-relativistic two-body 
problem is the effective-one-body, or EOB, approach. It allows one to model the 
full BBH evolution by including both perturbative results from the inspiral and 
ringdown phases as well as the full NR solution from the 
merger~\cite{Buonanno:1998gg,Albertini:2021tbt}. An alternative 
way to generate waveforms is to use phenomenological models to aid in detection and 
parameter inference. Although these models are fast, they use several
approximations which not only affect parameter estimation but also limit the
area of parameter space they can cover~\cite{Purrer:2019jcp,Estelles:2021jnz}.
To mitigate these issues, several groups have turned to the use of surrogate
modelling~\cite{Field:2013cfa,Varma:2018aht,Islam:2022laz,Walker:2022zob}. This type 
of modelling uses machine learning regression techniques informed by NR simulations. 
One must first build an accurate basis using the NR waveforms and then interpolate to 
construct new waveform templates in areas of parameter space not covered by 
simulations~\cite{Varma:2018mmi}. 

In this work, however, we take a different approach. We build a ringdown surrogate
model, \modelName, similar in methodology to~\cite{Varma:2018aht}.
We use Gaussian process regression (GPR), a supervised, machine learning 
algorithm, to interpolate between existing simulation data so that we can model a 
ringdown waveform at any point in parameter space. In other words, we do not need 
to directly model waveforms for all of the parameter space. Instead, we 
make a map from BBH system parameters like initial masses and spins to the
remnant parameters of mass, spin, and complex QNM mode amplitudes. This surrogate model 
allows us to make predictions for systems with parameters that have not been simulated.
More specifically, these systems have their spins aligned with the system's orbital angular 
momentum. We fit for the QNM amplitudes at a time $20 M$ after the peak of the $L^2$-norm 
of the waveform, or $u_0=u_{\mathrm{peak}} + 20M$, and use those to build our model.
Here, $u_0$ is defined as freely-specified start time of the model. The QNM fitting algorithm
is the same used in previous work~\cite{MaganaZertuche:2021syq}. 

The strength of our approach relies on the inclusion of multiple QNMs, memory, and the 
use of the correct Bondi-van der Burg-Metzner-Sachs (BMS) frame in our simulations, 
i.e., the superrest frame of the remnant BH~\cite{Bondi,Sachs,Boyle:2015nqa,Mitman:2021xkq}. The inclusion of memory and BMS frame fixing 
allows us to extract more accurate QNM amplitudes as well as more precisely model the 
ringdown portion of the waveform~\cite{MaganaZertuche:2021syq,Yoo:2023spi}. 

The LIGO-Virgo-KAGRA detector network has already observed ringdown-dominated 
signals, such as GW190521~\cite{LIGOScientific:2020iuh}. Soon, third-generation, 
ground-based detectors such as Cosmic Explorer (CE), the Einstein Telescope (ET), and 
space-based detectors such as the Laser Interferometer Space Antenna (LISA), 
will detect thousands of ringdown signals rich in QNM content~\cite{Sathyaprakash:2012jk,LIGOScientific:2016wof,
Holley-Bockelmann:2020lzp,Kalogera:2021bya,Baibhav:2019gxm}. 
Future detections require quick and robust modelling, which allows us to perform 
more accurate parameter estimation. However, existing surrogates perform poorly 
in the ringdown portion of a waveform. Therefore, we introduce 
aligned-spin ringdown surrogate \modelName. This new model is
publicly available through the python package
\texttt{surfinBH}~\cite{Varma:2018aht, vijay_varma_2018_1435832}.

\begin{figure}
  \centering
    \includegraphics[width=0.5\textwidth]{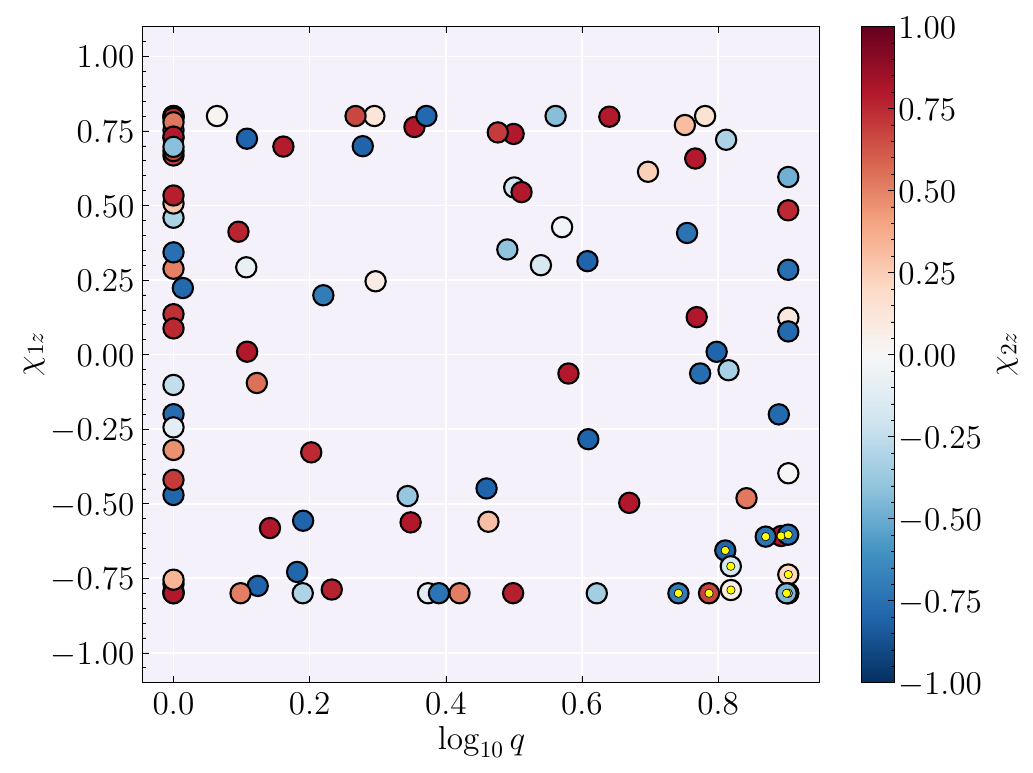}
  \caption{The parameter space spanned by the 102 NR simulations we use 
    to build\texttt{NRSur3dq8\_RD}. These are the same NR simulations used
    in surrogates \texttt{NRHybSur3dq8} and 
    \texttt{NRHybSur3dq8\_CCE}~\cite{Varma:2018mmi,Yoo:2023spi}.
    The axes show the spin on the heavier black hole versus the common
    logarithm of the mass ratio while the colorbar 
    indicates the spin on the lighter black hole. The yellow dots inside the 
    circles (bottom right corner) are those simulations with a larger retrograde $(2,2,0)$ mode 
    than the prograde mode.}
  \label{fig:initial_params}
\end{figure}

The paper is organized as follows. In Sec.~\ref{sec:conventions} we cover QNM conventions 
used throughout this work. In Sec.~\ref{sec:challenges} we discuss the main 
challenges of performing QNM amplitude fits. Then in Sec.~\ref{sec:surr_building}, 
we introduce Gaussian process regression as our machine learning algorithm 
of choice. We also present the waveform simulations used and the methods for 
training and testing the ringdown surrogate model. Finally, in 
Sec.~\ref{sec:results} we discuss model performance and results and comment 
on future work in Sec.~\ref{sec:conclusions}.

\section{QNM Conventions}\label{sec:conventions}

Each complex QNM frequency, $\omega$, is labelled by numbers $(\ell, m,n)$ where 
$(\ell,m)$ correspond to the angular numbers and $n=0,1,\ldots$ is an overtone 
number. Stable mode solutions live in the lower half-plane i.e., 
$\mathrm{Im}[\omega]<0$, and there is additional symmetry between the left and right 
half-planes such that $-\omega_{\ell, -m}^{*}$ is the mirror mode of 
$\omega_{\ell, m}$~\cite{Press:1973zz,London:2018nxs,Lim:2019xrb,Isi:2021iql,Li:2021wgz}.
Therefore, modes where $\mathrm{Re}[\omega]>0$ are known as ``ordinary'' and 
those with $\mathrm{Re}[\omega]<0$ are ``mirror'' modes. 

In addition to this symmetry, each QNM frequency has two solutions denoted by 
the superscript $p$: one solution that is co-rotating with the BH, or prograde 
($p=+1$), and another counter-rotating with the BH, called retrograde ($p=-1$). 
We call a mode prograde if
$\mathrm{sgn}(m)=+\mathrm{sgn}(\mathrm{Re}[\omega])$ and retrograde if
$\mathrm{sgn}(m)=-\mathrm{sgn}(\mathrm{Re}[\omega])$, following the naming 
convention of~\cite{MaganaZertuche:2021syq}. This
nomenclature breaks down for cases where $m=0$, where neither of the two 
solutions are known to be dominant.

Putting the above information together, we express a ringdown waveform as
\begin{align}
\label{eq:full-analytical}
h^{Q}(u,\theta,\phi)&=\sum_{\ell',m,n,p}\Bigg[ \mathcal{A}^{p}_{\ell' m n} e^{-i\omega^{p}_{\ell' m n}\left(u-u_{\text{ref}}\right)}\nonumber\\
&\phantom{=.\sum_{\ell' m n}[}\sum_{\ell} C_{\ell\ell'm}(a\omega^{p}_{\ell'mn}){}_{-2} Y_{\ell m}(\theta,\phi)\Bigg],
\end{align}
where the $\mathcal{A}^{p}_{\ell' mn}$ are complex amplitudes for each
QNM, $\omega^{p}_{\ell' m n}$
are the QNM frequencies, $C_{\ell \ell' m}(a\omega^{p}_{\ell'mn})$ are
called the spherical-spheroidal mixing coefficients in the conventions
of~\cite{Cook:2014cta,Stein:2019mop}, and
${}_{-2} Y_{\ell m}(\theta,\phi)$ are the spin-weight $-2$ spherical
harmonic functions. Here, $a=|J|/M$ is the BH spin parameter while
$\theta$ and $\phi$ represent the usual polar and azimuthal angles.
The QNM model vanishes for times earlier than the start time,
$h(u<u_{0})=0$.  The mode amplitudes are specified at a time
$u_{\text{ref}}$.  Almost everywhere we will choose
$u_{\text{ref}}=u_{0}$, except when testing for stability of mode amplitudes.

\section{Challenges in QNM Fitting}\label{sec:challenges}

Building a ringdown surrogate model requires at least a couple of important choices to
be made. The first is choosing a QNM model
start time $u_0$ for fitting the QNM amplitudes.
Since there is no clear time when the merger ends and ringdown begins, we cannot
prematurely choose the time when the waveform amplitude peaks to perform our fits. A
second challenge is to pick the number of modes that we want to include in the
surrogate model. Since different modes are important for different configurations of BBH systems,
it is more complex than choosing the 10 loudest modes of a single
simulation.

\subsection{On Choosing the Modes}\label{sec:choose_modes}

To answer these questions, we evaluate which QNM amplitudes are the most 
important in each simulation. Apart from aligning the peaks of the $L^2$ norm 
of the waveforms to be $u_{\mathrm{peak}}=0$ M, it is important to note
that there is rotational freedom around the $\hat{z}$-direction. 
Therefore, we can choose which $U(1)$ rotation to perform on these spin-aligned systems 
so that a better comparison can be made between simulations. Our choice is 
to make the $(\ell,m) = (2,2)$ spherical mode be real and positive at $u=u_{0}$.
We then proceed to fit for all 168 QNM amplitudes
$(\ell=2-4, |m|\leq\ell, n=0-7)$ for each simulation at times 
$u_0 - u_{\mathrm{peak}} \in [-10M,50M]$. This range in times allows us 
to evaluate how the absolute amplitudes evolve and which ones are the 
loudest in each simulation. It turns out that although the mode order varies 
between simulations, they have the most important modes in common. From this analysis, we also
notice that in a handful of the simulations, the amplitude of the retrograde mode is
dominant over the prograde mode amplitude. This remains true whether one
fits for all 168 modes or the 20 loudest modes. The simulations for which this occurs are 
denoted by a yellow dot on Fig.~\ref{fig:initial_params}. Through this figure, one
can see that the retrograde mode dominates in a certain area of parameter space, i.e., 
only for systems with a high mass ratio (q $\gtrsim 5.5$) and a high primary spin that is 
aligned with the $-\hat{z}$ direction, i.e., negatively spinning. 

\subsection{On Choosing the Fitting Times}

\begin{figure}
  \centering
    \includegraphics[width=\columnwidth]{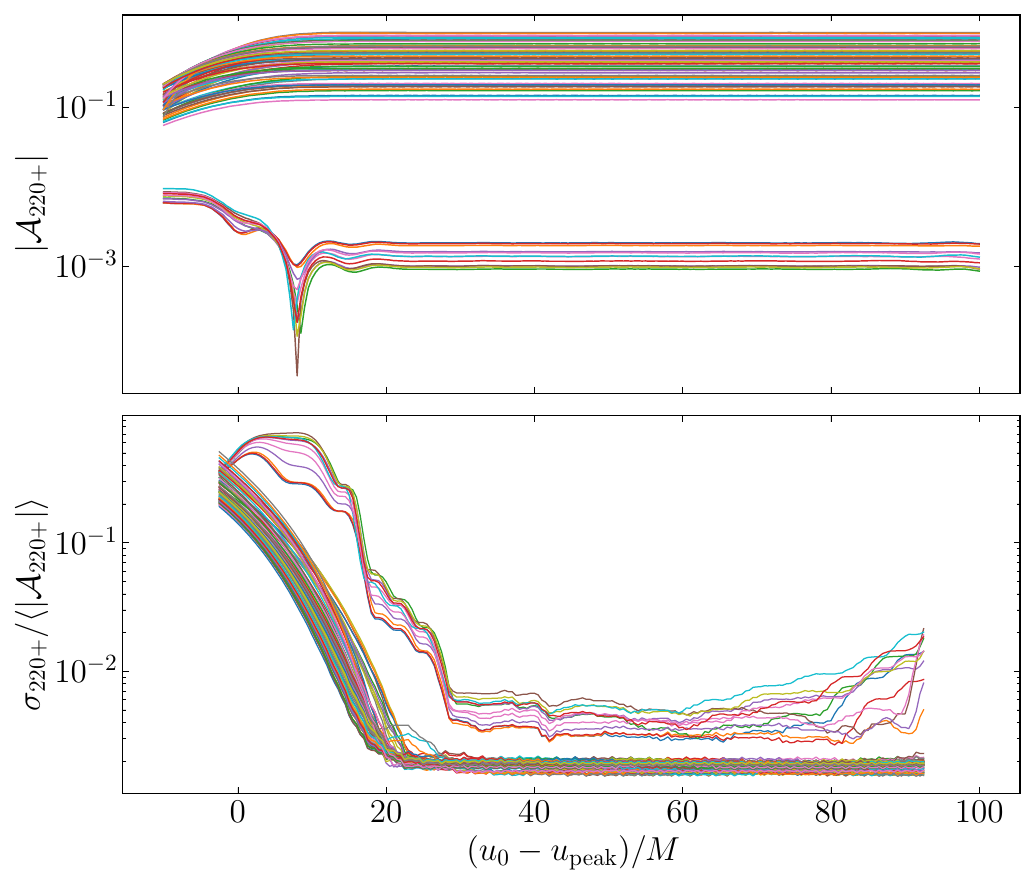}
  \caption{The top panel shows the absolute QNM amplitudes of the $(2,2,0,+)$
		mode as a function of the model start time $u_0$ for each simulation.
		The bottom panel shows the variance divided by the mean absolute amplitude over
    		a time window of $15M$ for this mode amplitude.
    }
  \label{fig:A220pm_abs_amp}
\end{figure}

To understand the fitting times, we constrain the number of modes we fit
to a set of 26 modes: $(2,\pm 2,0,\pm)$, $(2,\pm 2,1,\pm)$, $(2,\pm 1,0,\pm)$, 
$(2,0,0,\pm)$, $(3,\pm 3,0,\pm)$, $(3,\pm 2,0,\pm)$, and $(4,\pm 4,0,\pm)$. 
Unlike the conventions in~\cite{MaganaZertuche:2021syq}, here we define an 
individual mode to include only the prograde or retrograde solution such 
as $(\ell, |m|, n, +)$ or $(\ell, |m|, n, -)$, respectively. However, note that for building the surrogate mode we 
use a subset of the modes above, as seen in Sec.~\ref{sec:modes_modelling}. 
We fit the QNM amplitudes for times $u_0-u_{\mathrm{peak}} \in [-10M,100M]$ 
to evaluate which time or range in time is the most reliable for extracting QNMs. A 
physically motivated way to answer this question is by inspecting the stability of each 
mode as a function of the start time for each simulation. Specifically, we look at how the
absolute values of the QNM amplitudes evolve based on a chosen fitting start
time. If the amplitude of a mode is highly varying, we call this mode unstable, and 
therefore, we cannot fully rely on its value. To quantify this highly varying behavior, 
we find the variation in the mode amplitude by calculating its variance over a window
of time that avoids any bias. The variance of the QNM amplitude is given by
\begin{multline}
\label{eq:variance}
\sigma^2_{|\mathcal{A}_{\ell mn}|}(u_{0}) = \\
\frac{1}{\Delta u}\int^{u_0-\Delta u/2}_{u_0+\Delta u/2}
        \Big(|\mathcal{A}_{\ell mn}(u_0)| - \langle|\mathcal{A}_{\ell mn}|\rangle\Big)^2 du_0,
\end{multline}
where $\langle|\mathcal{A}_{\ell mn}|\rangle$ is the average of the
magnitude of the QNM amplitude over the window of time $\Delta u = 15 M$.
Since we are comparing fits with different choices of $u_{0}$, we
reference them all to the same fixed $u_{\text{ref}}$, taking out the
exponential decay.
For maximum stability, we choose a time within the window where the variation
is at a minimum. Note that a window that is too small will have almost no
variation while a large window may include values that are unphysical --- leading
to a high variation, as seen in the top panel of Fig.~\ref{fig:A220pm_abs_amp}.

\begin{figure}
  \centering
    \includegraphics[width=\columnwidth]{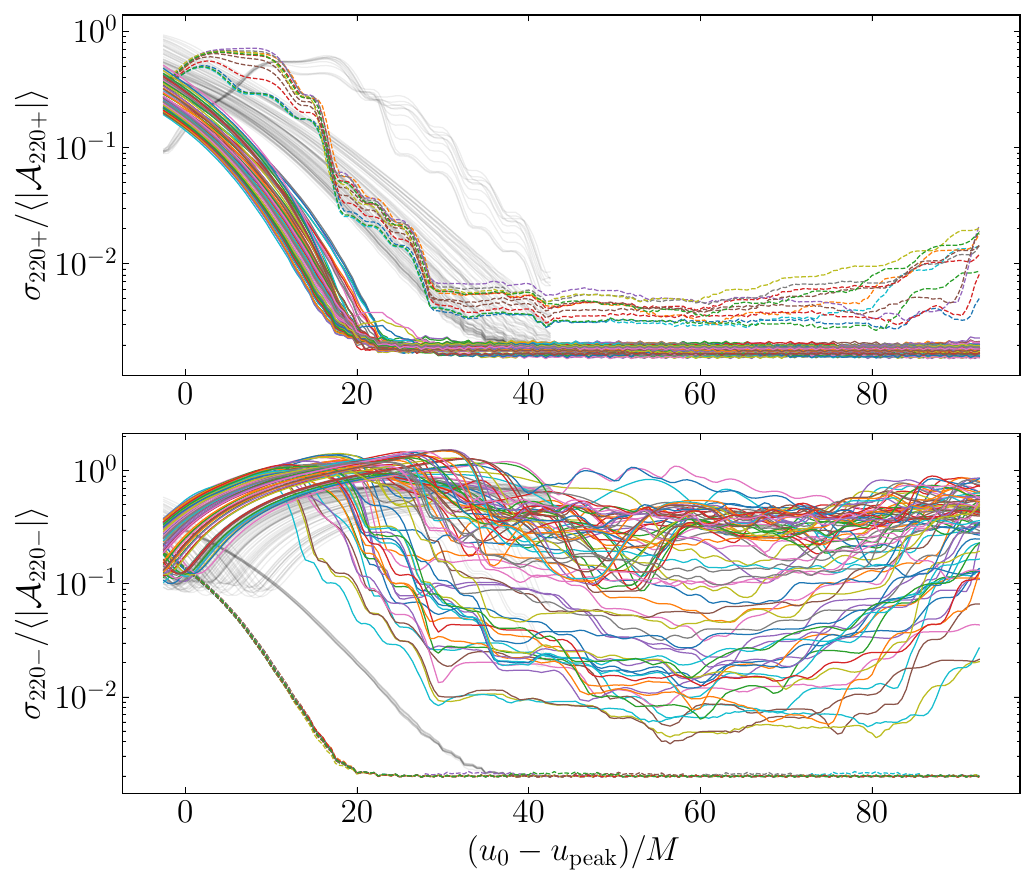}
    \caption{The top panel shows the variance divided by the mean absolute
    amplitude over a $15M$ time window of the $(2,2,0,+)$ mode as a function of 
    the model start time $u_0$ for each simulation. The bottom panel shows the 
    corresponding quantities for the $(2,2,0,-)$ mode. The color lines
    represent the extraction of the modes when the $(2,2,1,\pm)$ mode is
    included in the fitting process while the light gray lines in the
    background represents the fits with only the fundamental modes are included.
    }
  \label{fig:variances}
\end{figure}

We are interested in the later times when the behavior seems to have stabilized 
and the modes can be reliably extracted. Note that some simulations show a 
smaller absolute amplitude on the order of $10^{-3}$. These are the BBH systems 
for which the retrograde mode is dominant over the prograde mode and, therefore, 
the prograde mode has a smaller amplitude. To get a better idea of the $(2,2,0,+)$ 
mode stability we show in the bottom panel the variance over the mean of the 
absolute amplitudes over a $15M$ time window. For the $(2,2,0,+)$ mode, it is clear 
that times around and after $u_0-u_{\mathrm{peak}} = 20M$ are more stable and, 
thus, more reliable to use in our fits. If we fit for the $(2,2,0,-)$ mode, the curves 
reverse, i.e., the systems with a dominant retrograde mode have low variations after 
$u_0-u_{\mathrm{peak}} = 20M$ while the prograde-dominant systems show a 
jagged line. This can be seen by comparing the colored curved on the top and 
bottom panels of Fig.~\ref{fig:variances}.

An important finding to note is that the variance over the mean 
absolute amplitude decreases when fitting the additional $(2,2,1,\pm)$
overtone If we do not include the first overtone, our amplitude 
variances would look like the gray lines shown in Fig.~\ref{fig:variances}. The top
panel shows how the inclusion of the first overtone brings mode stability closer 
to the peak of the waveform. This tells us that including the first overtone, 
which can be very loud in several simulations, improves our fits. This is no 
surprise since similar behavior can be seen in the mismatch plots when including 
more than just the fundamental overtone~\cite{Giesler:2019uxc}. However, 
one must be careful when including overtones as this can produce overfitting. 
Overtones higher than $n = 2$ contribute very little power at late times like 
$u_0-u_{\mathrm{peak}} = 20 M$ since they decay rapidly.

As mentioned in the discussion above, 
the variance normalized by the mean absolute amplitude of $(2,2,0,-)$ is just
as low in retrograde-dominated simulations as for the prograde amplitude in
the prograde-dominated simulations. A striking feature of this plot is also
the high variance in the retrograde mode at all times. This is the first
telltale sign that fitting for these amplitudes quickly becomes challenging.

In addition to varying the fitting start time, we can use the mismatch between 
the NR waveform and our model at different start times to get a more physically 
motivated idea of which time choice to make. We define mismatch as
\begin{align}
\label{eq:mismatch}
\mathcal{M}(a,b)\equiv1 - \text{Re}\left[\frac{\langle a, b \rangle}{\sqrt{\langle a, a \rangle\langle b, b \rangle}}\right],
\end{align}
where $\langle a,b\rangle$ is the inner product between waveforms $a$ and $b$ defined as 

\begin{align}
\label{eq:decomp-inner-product}
\langle a,b\rangle
&= \int_{S^{2}} \int_{u_0}^{u_f} a^{*}(u,\theta,\phi) b(u,\theta,\phi) \, \rmd u \rmd \Omega \\
&=\int_{u_0}^{u_f} \sum_{\ell,m} a^{*}_{\ell m}(u) b_{\ell m}(u) \, \rmd u
\end{align}
and $[u_0, u_f]$ is the interval of time for fitting the waveform with a QNM model.
As seen in Fig.~4 of~\cite{MaganaZertuche:2021syq}, $u_0-u_{\mathrm{peak}} = 20 M$ 
is a time when the mismatch reaches a minimum but is still above numerical error. 
In fact, regardless of the number of modes one chooses to model, the mismatch 
minimum dances around $u_0-u_{\mathrm{peak}} = 20 M$. Therefore, we choose 
to fit all the QNM amplitudes at this same time, i.e., the time at which we specify 
the QNM amplitudes is same for all QNMs.

\section{Building the Surrogate}\label{sec:surr_building}

There are various ways to approach building a ringdown surrogate. In this work, we
do not need to directly model the waveforms' time dependence for all of the parameter space but
instead, we build a map from BBH system parameters like initial masses and spins 
to the remnant parameters of mass, spin, and complex-valued QNM mode amplitudes. That is, we use 
all of the BBH system data and make predictions for systems with parameters that 
have not been simulated.

The surrogate model is implemented using Gaussian process regression, which 
provides us with a sophisticated mathematical tool to find joint Gaussian
distributions between the system data and the predictions~\cite{Rasmussen2006}. 
This method allows us to interpolate between data sets and estimate the
errors of the fitted quantities. More specifically, we are
using the GPR fitting method employed in~\cite{Varma:2018aht}, which is built using
\texttt{scikit-learn}~\cite{DBLP:journals/corr/abs-1201-0490}.

In the rest of this section, we discuss the choice of modes included in the ringdown
surrogate model. We also show the preparation needed to curate a dataset for
GPR as well as the methods used for training and testing the dataset.

\subsection{Modes Modelled}\label{sec:modes_modelling}

Based on the analyses performed in Sec.~\ref{sec:challenges}, we can confidently
include 10 QNM amplitudes in this ringdown surrogate model:
$(2,2,0,+)$, $(2,-2,0,-)$, $(2,2,1,+)$, $(2,-2,1,-)$, 
$(2,0,0,\pm)$, $(3,2,0,+)$, $(3,-2,0,-)$, $(4,4,0,+)$, 
and $(4,-4,0,-)$. Notably, due to the large variance of the retrograde modes across 
simulations (see Fig.~\ref{fig:variances}), our surrogate model does not include 
them as the resulting GPR fits were inaccurate.
As we have seen from Fig.~\ref{fig:initial_params}, this affects a 
corner of parameter space with a high mass ratio and highly negative effective
spin,
\begin{align}
\label{eq:chi_eff}
\chi_{\mathrm{eff}} = \frac{q\chi_1 + \chi_2}{1+q} = \frac{m_1\chi_1 + m_2\chi_2}{m_1+m_2} .
\end{align}
One possible remedy may be to increase the weight that GPR gives to the
retrograde modes for these simulations. Another possible option would be to
``zero out'' the amplitude values that are not dominant, such that the
prograde(retrograde) mode would be non-zero for simulations where the
prograde(retrograde) is dominant over the retrograde(prograde). The downside to
this solution is that we may miss out on important physics. In future work, we will 
be investigating the best way to include retrograde modes. 

A second noticeable aspect of the surrogate model is that we do not take into account the 
$(2,\pm1,0,+)$ and $(3,\pm3,0,+)$ modes. Not very surprisingly, it
turns out that the QNM amplitudes for these modes are very small when looking at
BBH systems with a mass ratio of one. The inherent symmetries of equal mass 
systems cause multipoles with an odd $m$ number to be zero. Since many of the
simulations have a mass ratio of one or close to one, GPR is not able to predict as 
smoothly across parameter space. One can in
principle include the modes in the model, but their trustworthiness would be
questionable. As of now, we do not have a clear way to remedy this challenge,
but it is something we will be working toward shortly as we improve
the robustness of the surrogate model.

\subsection{Numerical Relativity Waveforms and Dataset}

For this surrogate model, we use 102 binary black hole mergers evolved by 
the SXS Spectral Einstein Code (SpEC)~\cite{SpECCode}. These are the same 
systems used to create \texttt{NRHybSur3dq8} and 
\texttt{NRHybSur3dq8\_CCE}~\cite{Varma:2018mmi,Yoo:2023spi}. The waveforms are extracted using 
Cauchy-characteristic evolution (CCE) from SpECTRE and a BMS 
transformation is applied to them. In this way, they are supertranslated so that 
the remnant black hole is in the superrest frame~\cite{Moxon:2020gha,Moxon:2021gbv,CodeSpECTRE,Mitman:2020pbt,Mitman:2020bjf,Mitman:2021xkq}. 
The mass ratios spanned by this set of simulations are in the range 
$q \leq 8$ with the primary and secondary spins ranging as 
$\chi_{z}^{1,2} \in [-0.8, 0.8]$, as shown in Fig.~\ref{fig:initial_params}.
The initial parameters of these simulations are specified at a reference time after 
the effects of junk radiation have sufficiently decayed~\cite{Boyle:2019kee}.

The dataset consists of initial BBH and remnant parameters for each BBH
system. The initial parameters we use are $\{\log(q), \hat{\chi},
\chi_a\}$, where $\hat{\chi}$ is the spin parameter that enters the GW phase at
leading order~\cite{Khan:2015jqa,Ajith:2011ec,Cutler:1994ys,Poisson:1995ef} in 
the post-Newtonian expression and is defined as
\begin{align}
\label{eq:chi_hat}
\hat{\chi} = \frac{\chi_{\mathrm{eff}} - 38\eta (\chi_{1z}+\chi_{2z})/113}{1-76\eta/113},
\end{align}
and $\chi_a$ is the anti-symmetric spin given by
\begin{align}
\label{eq:chi_a}
\chi_a = \frac{1}{2}(\chi_{1z} - \chi_{2z}).
\end{align}
Just as in~\cite{Varma:2018aht}, we found this parametrization gives more accurate
predictions than using $\{q, \chi_{1z}, \chi_{2z} \}$. We obtain the remnant mass 
and spin from $\scri^{+}$ by using Poincaré charges rather than the apparent
horizon~\cite{Iozzo:2021vnq}. 

As done in Sec.~\ref{sec:choose_modes}, we begin preparing our
dataset by first aligning the waveforms so that the peak of the $L^2$ norm is at 
$u_0 - u_{\mathrm{peak}} = 0 M$. As before we fix the $U(1)$ rotational symmetry
of the systems so that the $(2,2)$ spherical mode is real and positive at
$u_0 - u_{\mathrm{peak}} = 20 M$. 
Once we have rotated our waveforms, we are ready to use them in the
QNM fitting algorithm~\cite{scri_url}. Fitting for the QNMs at $u_0 - u_{\mathrm{peak}} = 20 M$
gives us the QNM amplitude values that we will be using in our training and
testing dataset.
Since \texttt{scikit-learn}'s GPR does not take complex values,
we decompose our
complex QNM amplitudes into real and imaginary parts. Therefore, 
we model $\{m_{f}, \chi_{f}, \Re[\mathcal{A}_{\ell mn}], \Im[\mathcal{A}_{\ell mn}]\}$ 
as a function of $(\log(q), \hat{\chi},
\chi_a)$. For example, our remnant mass model is given by $m_{f}(\log(q), \hat{\chi}, \chi_a)$.
Machine learning (ML) algorithms require the data to be standardized, i.e., have a
zero mean and unit variance. We pre-process the data before training the
algorithm as in~\cite{Varma:2018aht}, i.e., we subtract a linear fit and the
mean from the data points and then normalize them. When evaluating the fits,
the inverse transformations are applied.

Before analyzing the results, we need to calculate the numerical error in our 
waveforms. We perform this calculation by taking the highest and second-highest 
resolution waveforms such that the error between them is given by
\begin{equation}
\label{eq:errors}
\mathcal{E}[a,b]\equiv\frac{1}{2} \frac{\langle a-b | a-b\rangle}
{\langle a | a\rangle},
\end{equation}
where $a$ represents the highest resolution NR waveform available and $b$ is the second
highest resolution waveform. This equation was first introduced 
in~\cite{Blackman:2017dfb} as related to the weighted average of the all-mode
mismatch.
\subsection{Training and Testing}
The most common way to train any ML algorithm is to separate the
dataset into two parts: training and testing.
Often, $70 \%$ of a dataset is used in training and $30 \%$ for testing. However, this
is usually for datasets that contain tens of thousands of data points or more.
For small datasets like ours, this is a problem since only a fraction of the data would
be used in training. To mitigate this problem, we instead employ k-fold cross-validation.
This method entails equally dividing a dataset into $k$ smaller
subsets, using one of these subsets for predicting and taking the remaining
$k-1$ for training. This is then repeated $k$ times until each subset is used
for validation, 
For large datasets, this could
present a problem since one must train $k$ models instead of one. However, our
dataset is small enough that the additional training time does not outweigh the benefits
of the method. We use a value of $k = 20$, where all datasets contain $5$
simulations except for three of them which contain $6$ since the number of 
simulations is not divisible by 5. This way of splitting the
dataset allows us to train and test over our entire dataset.
Additionally, k-fold cross-validation gives us errors for each k-fold we
test. Since we use a value of $k = 20$, we get $20$ values of testing errors,
which we average and use as an estimation of the model's error.

\section{Results}\label{sec:results}
\begin{figure}
    \centering
    \includegraphics[width=\columnwidth]{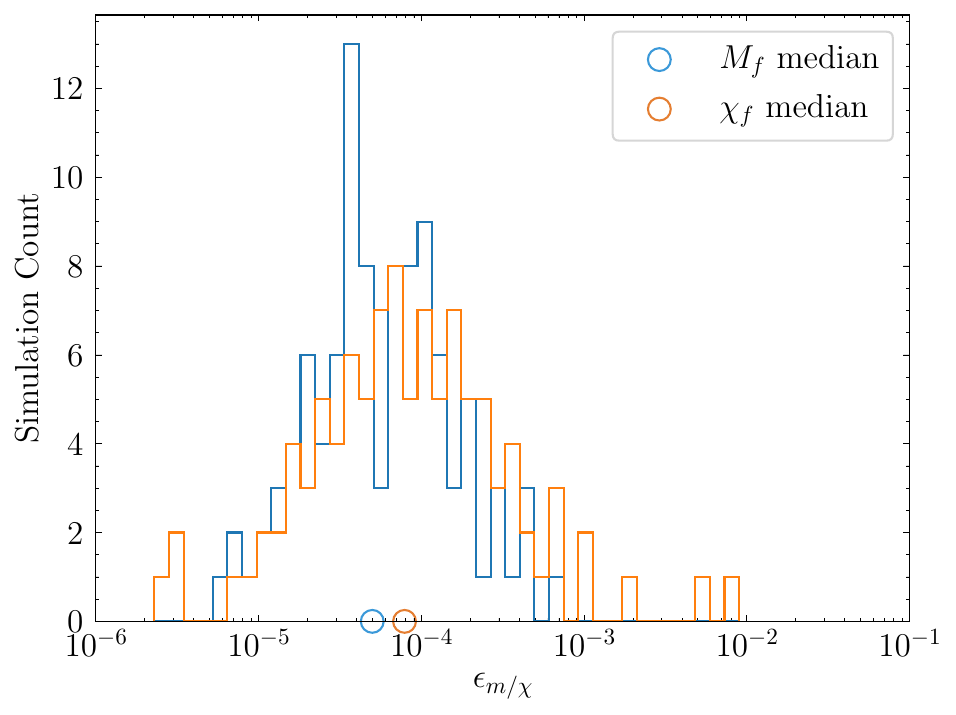}
    \caption{A histogram showing the relative error in remnant mass (blue) and the
    absolute difference in the remnant spin (orange) as calculated in Eqs.~\eqref{eq:mass_error}
    and~\eqref{eq:spin_error}, respectively. The median errors are denoted by the circles on the horizontal axis.
        }
  \label{fig:spin_chi_errors}
\end{figure}

Running GPR with the settings discussed above, we get predictions for each
k-fold testing dataset. Figure~\ref{fig:spin_chi_errors} shows how well the surrogate model
does in recovering the remnant mass and spin. The error in the remnant 
mass is the relative error defined as 
\begin{align}
\label{eq:mass_error}
    \epsilon_m = \frac{|m^{\mathrm{NR}}_f -
    m^{\mathrm{surr}}_f|}{m^{\mathrm{NR}}_f} ,
\end{align}
while the error in spin is the absolute difference given by
\begin{align}
\label{eq:spin_error}
    \epsilon_{\chi} =|\chi^{\mathrm{NR}}_f - \chi^{\mathrm{surr}}_f|.
\end{align}
Note that both errors are low and agree with the results from the remnant surrogate
of~\cite{Varma:2018aht}. The median error in the remnant mass is
$5\times 10^{-5}$ while the median error in the remnant spin is $\sim 8 \times
10^{-5}$. The 95th percentile errors are $3.2 \times 10^{-3}$ and $6.8 \times
10^{-4}$, respectively.

\begin{figure}
  \centering
    \includegraphics[width=\columnwidth]{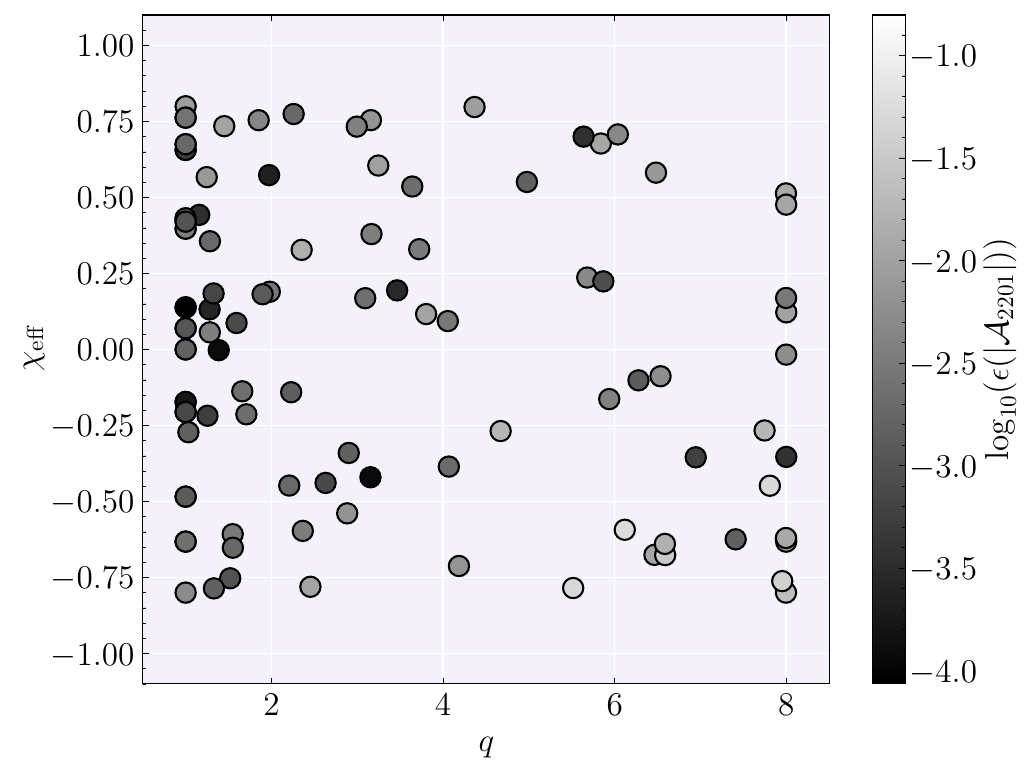}
    \caption{The relative errors of the magnitude of the $(2,2,0,+)$ mode amplitude
    across parameter space.
        }
  \label{fig:2201_amp_errs}
\end{figure}

\begin{figure}
  \centering
    \includegraphics[width=\columnwidth]{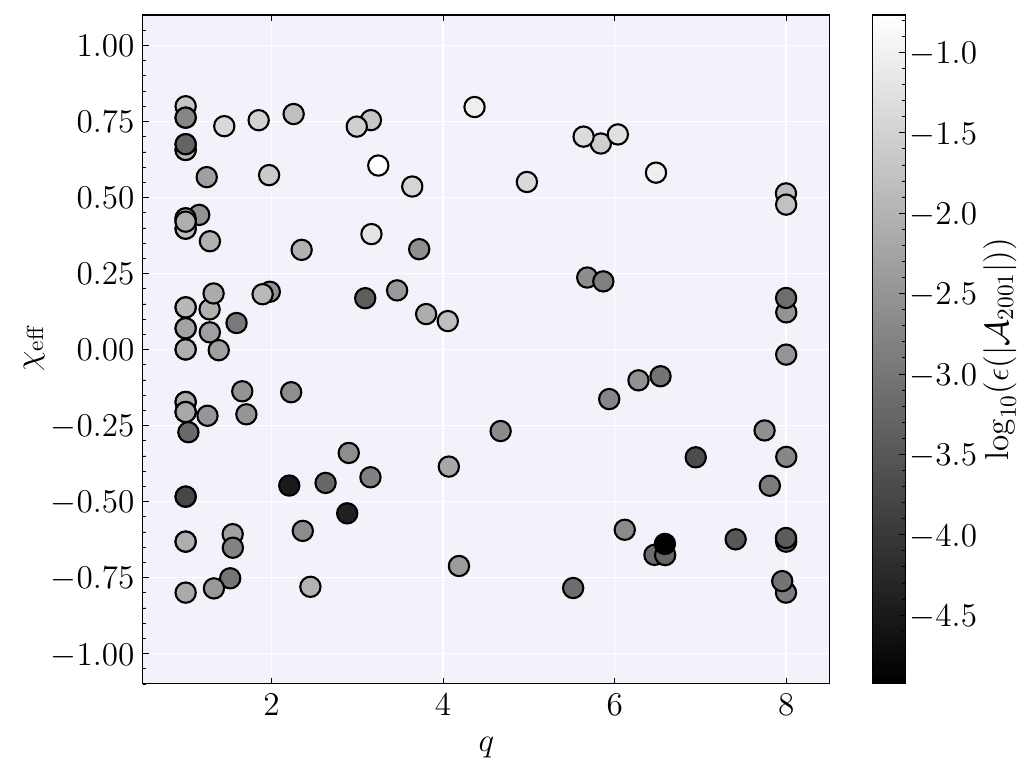}
  \caption{The relative errors of the magnitude of the $(2,0,0,+)$ mode amplitude
    across parameter space.
        }
  \label{fig:2001_amp_errs}
\end{figure}

\begin{figure}
  \centering
    \includegraphics[width=\columnwidth]{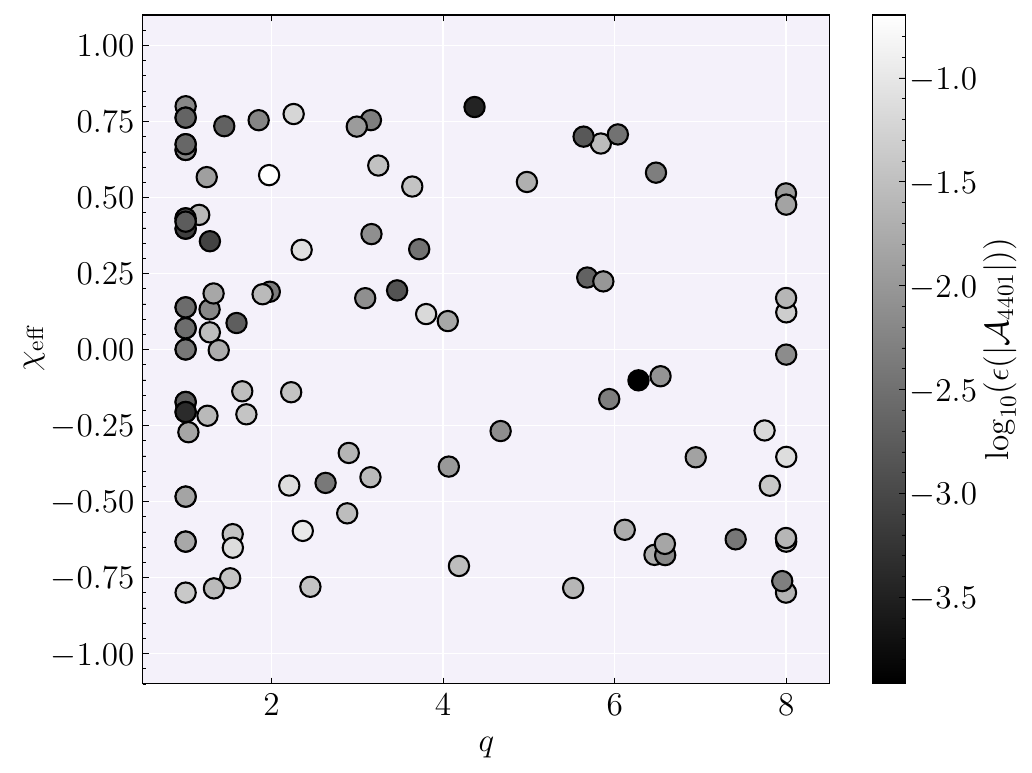}
  \caption{The relative errors of the magnitude of the $(4,4,0,+)$ mode amplitude
    across parameter space.
        }
  \label{fig:4401_amp_errs}
\end{figure}

For errors in predicting the QNM amplitudes, we calculate the
relative errors,
\begin{align}
  \label{eq:absamp_rel_error}
      \epsilon(|\mathcal{A}_{\ell mn}|) = 
      \frac{|(|\mathcal{A}_{\ell mn}^{\mathrm{QNM}}| - |\mathcal{A}_{\ell
      mn}^{\mathrm{surr}}|)|}
      {|\mathcal{A}_{\ell mn}^{\mathrm{QNM}}|},
  \end{align}
of the magnitude $|\mathcal{A}_{\ell mn}|$ over 
parameter space. Here $\mathcal{A}_{\ell mn}^{\mathrm{QNM}}$ denotes the complex-valued QNM amplitudes extracted from the NR data while $\mathcal{A}_{\ell mn}^{\mathrm{surr}}$ denotes the surrogate model of $\mathcal{A}_{\ell mn}^{\mathrm{QNM}}$. Fig.~\ref{fig:2201_amp_errs} illustrates the error in predicting the $(2,2,0,+)$ mode
amplitude over parameter space. Furthermore, we show the errors for the $(2,0,0,+)$ and $(4,4,0,+)$ modes 
in Figs.~\ref{fig:2001_amp_errs} and~\ref{fig:4401_amp_errs}.
Overall, the median values for the amplitude errors are listed in
Table~\ref{tab:error_table}. Although the table only lists the $+m$ prograde modes,
the relative errors for the $-m$ prograde modes are the same due to mode symmetries 
in spin-aligned systems. The first two
error columns show the values for the relative amplitude errors for the real and 
imaginary components, calculated as
\begin{align}
\label{eq:real_rel_error}
    \epsilon(\Re[\mathcal{A}_{\ell mn}]) = \left|
    \frac{\Re[\mathcal{A}_{\ell mn}^{\mathrm{QNM}}] - \Re[\mathcal{A}_{\ell
    mn}^{\mathrm{surr}}]}
    {\Re[\mathcal{A}_{\ell mn}^{\mathrm{QNM}}]}\right|.
\end{align}

For the last column, we combine the real
and imaginary components to get the complex QNM amplitude and calculate the 
relative error of the absolute amplitudes using Eq.~\eqref{eq:absamp_rel_error}.
Unsurprisingly, the absolute amplitude errors are smallest for the 
$(2,2,0,+)$ and $(2,-2,0,-)$ modes which are always the dominant modes in these 
simulations. Nevertheless, one can see that even modes with smaller magnitudes 
are well recovered.
\begin{table}[]
\centering
  \sisetup{table-format = 1.4, table-alignment-mode = format}
  \begin{tabular}{c|
    S[table-number-alignment = center]|
    S[table-number-alignment = center]|
    S[table-number-alignment = center]}
\hline\hline\noalign{\smallskip}
QNM      & {$\epsilon(\Re[{\mathcal{A}_{\ell mn}}])$} & {$\epsilon(\Im[{\mathcal{A}_{\ell mn}}])$} & {$\epsilon(|\mathcal{A}_{\ell mn}|)$} \\ \hline
(2,2,0,+) & 0.0020            & 0.061  & 0.0024  \\
(2,2,1,+) & 0.060             & 0.064  & 0.050   \\
(2,0,0,+) & 0.0026            & 0.0089 & 0.0037  \\
(3,2,0,+) & 0.024             & 0.012  & 0.0084  \\
(4,4,0,+) & 0.035             & 0.011  & 0.011   \\
\noalign{\smallskip}\hline\hline
\end{tabular}
\caption{Table showing the median relative errors of the real, imaginary, and absolute 
    QNM amplitudes predicted. The relative errors for each simulation are defined
    in Eqs.~\eqref{eq:absamp_rel_error} and~\eqref{eq:real_rel_error}.}
\label{tab:error_table}
\end{table}

\begin{figure}
  \centering
    \includegraphics[width=\columnwidth]{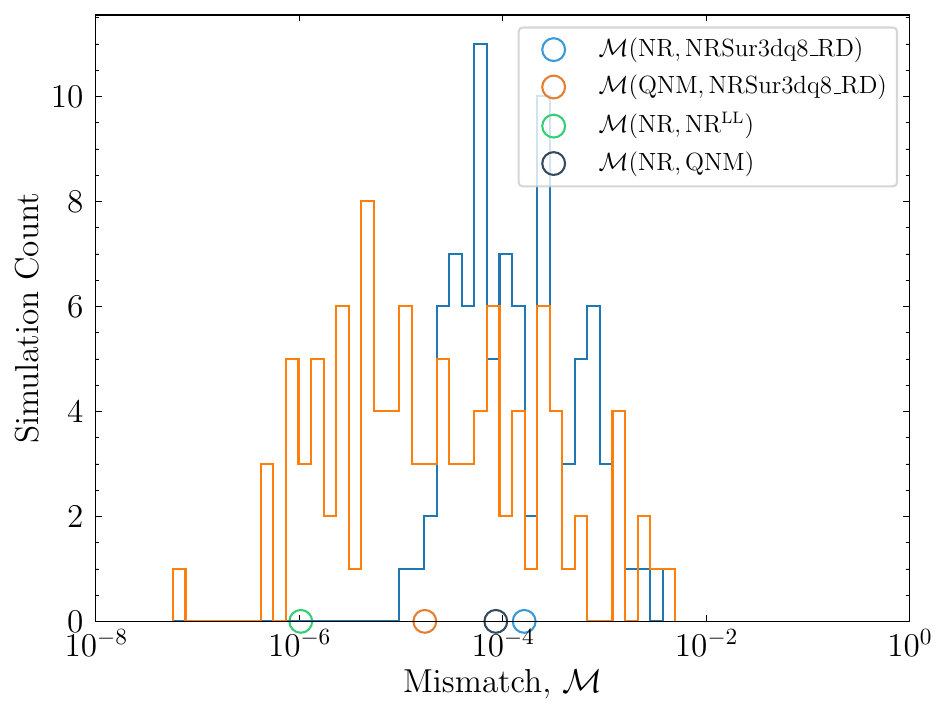}
  \caption{The mismatch distribution of simulations. The blue histogram shows the 
  		mismatch between the NR waveforms and surrogate model, while the 
		orange histogram shows the mismatch between the QNM and surrogate
		models. The green circle on the horizontal axis highlights the median numerical 
		error with the label ``LL" for ``Lower Lev", i.e., the next-highest resolution 
		simulation. The median mismatch between NR and the QNM model is shown 
		with the black circle.
        }
  \label{fig:NR_surr_mismatches}
\end{figure}

In Fig.~\ref{fig:NR_surr_mismatches} we show the mismatches
between the different types of waveforms for all the simulations. The
calculation of the mismatches includes only the QNMs in the surrogate and is
not for the true all-mode mismatches. The blue histogram shows the mismatch
distribution between NR and the surrogate waveform. The median value is of 
$1.6 \times 10^{-4}$ and can be seen in blue on the horizontal axis. The orange 
histogram shows the mismatch distribution between the QNM model and the surrogate 
model. As expected, the median mismatch, shown in orange, is lower at 
$1.7 \times 10^{-5}$. We also include the median values for
the mismatches between NR and the QNM model in black, which is just slightly lower than
that of the surrogate model, and the mismatch from the numerical waveform
resolutions in green. The mismatch in the highest and the next-highest waveform
resolutions is around $1 \times 10^{-6}$, as calculated from Eq.~\ref{eq:errors}. Since our 
model mismatches are higher than those coming from numerical error, we do not 
expect any overfitting.

\begin{figure*}
   \centering
   \includegraphics[width=\linewidth]{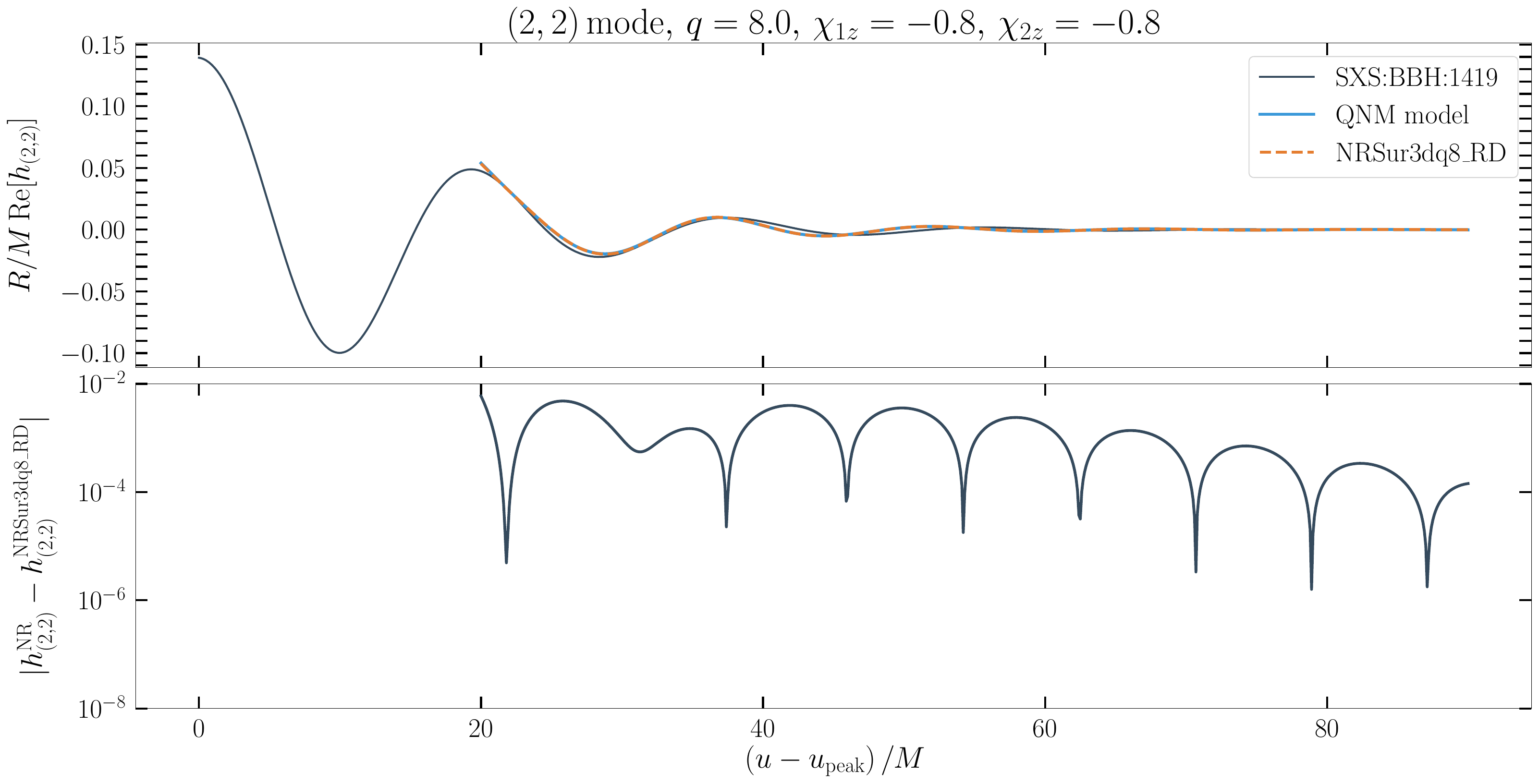}
   \caption{%
        We plot the worst performance, or greatest mismatch, between the real 
        component of the $(2,2)$ mode of a CCE waveform, the best-fit QNM model built
        from the $(2,2)$ mode with $n=0$ and the first overtone, and
        the surrogate. The upper panel shows NR
        waveform \texttt{SXS:BBH:1419} (black), the QNM model (blue), and \modelName{} 
        (orange). The lower panels show the residuals between NR and
        \modelName{} waveforms. 
        }
    \label{fig:strain220max}
\end{figure*}

\begin{figure*}
   \centering
    \includegraphics[width=\linewidth]{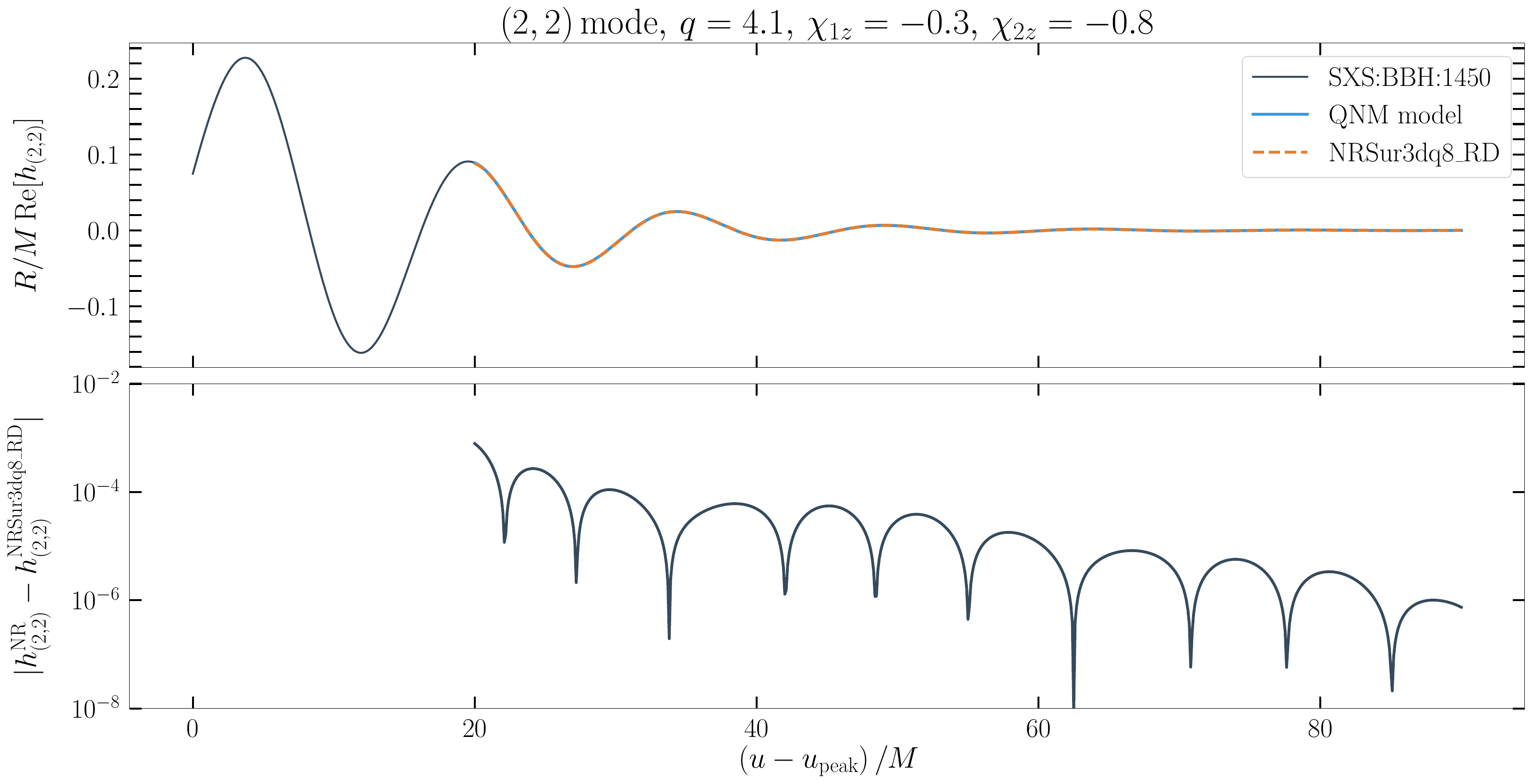}
    \caption{%
        Similar to Fig.~\ref{fig:strain220max}, but the case with smallest
        mismatch between NR waveform \texttt{SXS:BBH:1450} and the surrogate.
        }
    \label{fig:strain220min}
\end{figure*}

Finally, Figs.~\ref{fig:strain220max} and~\ref{fig:strain220min} show how these 
predicted mode amplitudes translate to the
waveform strain. In
both plots, the top panel shows the real part of the $\ell = m = 2$ NR waveform strain (black), the QNM model (blue), and the surrogate model (orange) for two 
different systems. Both the QNM model and the surrogate model include the 
$(2,2,0,+)$ and $(2,2,1,+)$ modes. In Fig.~\ref{fig:strain220max}, we take the
waveform that has the largest mismatch between NR and our QNM model while in
Fig.~\ref{fig:strain220min} we choose the NR waveform with the smallest mismatch.

\begin{figure*}
  \centering
    \includegraphics[width=\linewidth]{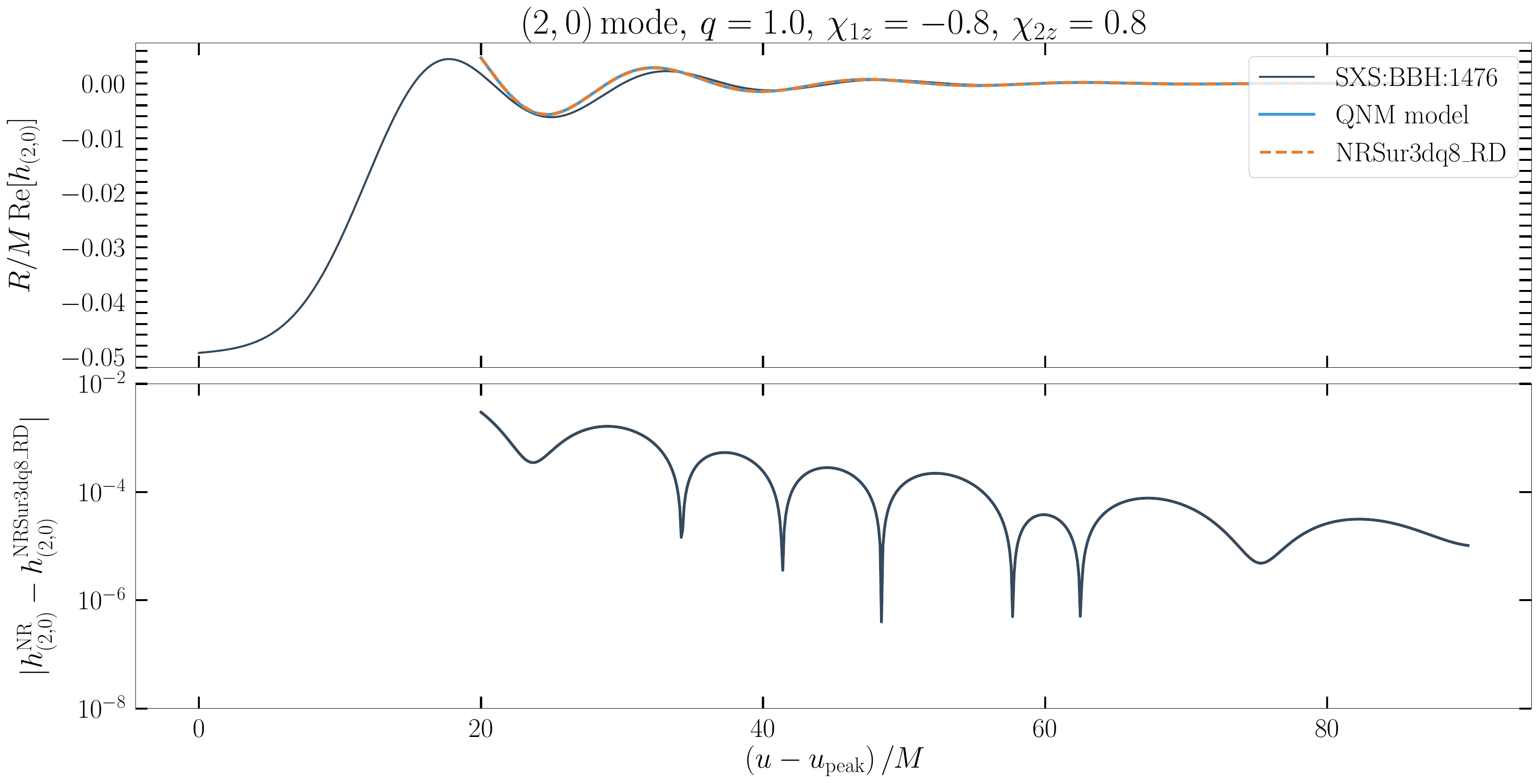}
  \caption{%
        Similar to Fig.~\ref{fig:strain220max}, but the case with median
        mismatch between NR waveform \texttt{SXS:BBH:1476} and the surrogate,
        showing the $(2,0)$ mode and the reconstruction using both
        prograde and retrograde modes $(2,0,0,\pm)$.
        }
  \label{fig:strain200med}
\end{figure*}

\begin{figure*}
  \centering
    \includegraphics[width=\linewidth]{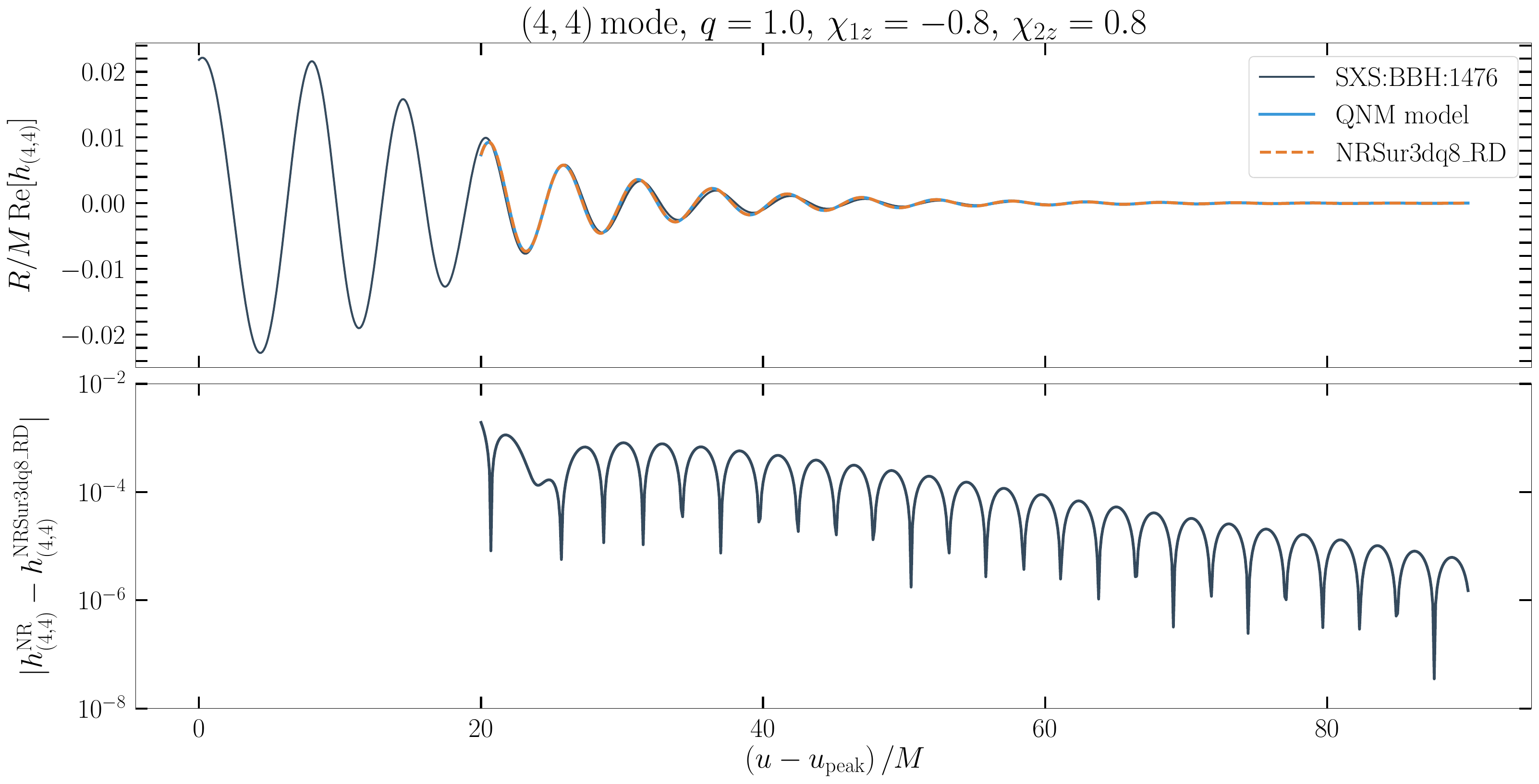}
  \caption{%
        Same NR waveform as Fig.~\ref{fig:strain200med}, but showing the real
        component of the $(4,4)$ mode using the fundamental mode.
        }
    \label{fig:strain440med}
\end{figure*}

In the bottom panels of both figures, we show the residuals between NR and the
surrogate model. Focusing on the black line on Fig.~\ref{fig:strain220max}, we see that at
$u - u_{\mathrm{peak}} = 20 M$ the residual is on the order of $10^{-3}$. 
This is the worst-case scenario we expect to predict
within the training region. Not surprisingly, the waveform with the worst mismatch
corresponds to a simulation with a mass ratio of $q=8$ and spins of $\chi=-0.8$. 
This is the area of parameter space that showed a preference for the retrograde 
mode amplitudes (which is neglected in our both our QNM and surrogate models), and therefore, 
the $(2,2)$ mismatch is higher. On the other extreme, Fig.~\ref{fig:strain220min} 
shows a residual of $10^{-4}$. We also show 
$10^{-3}$ order of magnitude residuals of the surrogate model for the 
$(2,0,0,\pm)$ and $(4,4,0,+)$ modes in Figs.~\ref{fig:strain200med}
and~\ref{fig:strain440med}, respectively. From 
Figs.~\ref{fig:strain220max},~\ref{fig:strain200med}, and~\ref{fig:strain440med} one can 
visually see the differences between the NR waveform and \modelName{}. This 
highlights the importance of using more than just the $(2,2)$ for modeling a BBH 
system~\cite{Giesler:2019uxc,Cook:2020otn,London:2014cma,Baibhav:2018rfk,Berti:2014fga,Dhani:2021vac,Finch:2021iip,Dhani:2020nik,Li:2021wgz}. 


\begin{figure*}
  \centering
    \includegraphics[width=\linewidth]{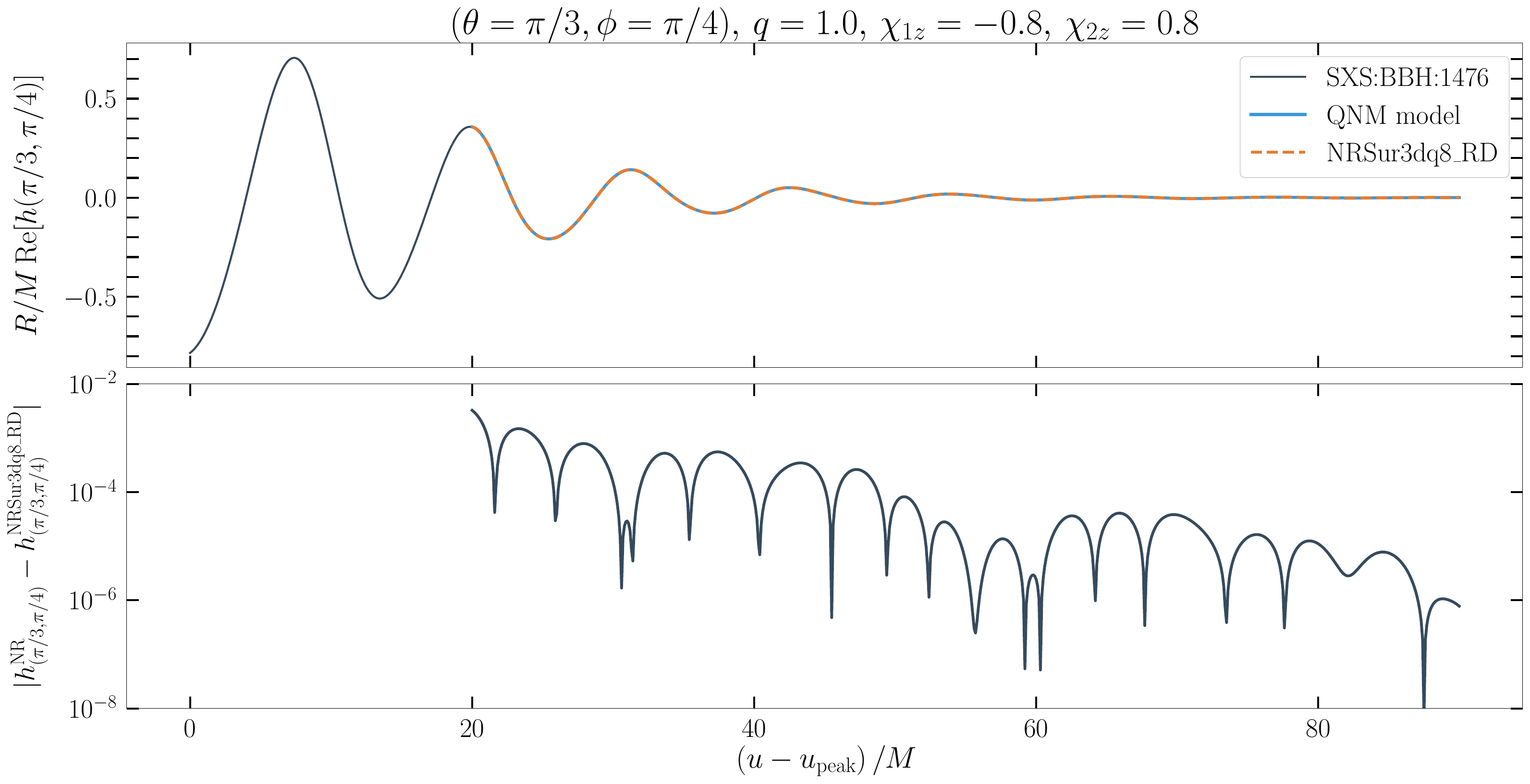}
    \caption{%
        Same NR waveform as Fig.~\ref{fig:strain200med}. The real
        component of the strain measured by an observer at $(\theta =
        \pi/3 , \phi = \pi/4)$.  The strain is reconstructed from all
        harmonics used in \modelName.
        }
  \label{fig:strainlmmed}
\end{figure*}

In Fig.~\ref{fig:strainlmmed}, we show the full modelling power of the surrogate
model by using the NR waveform containing only the modes modelled by the surrogate 
overlaid with the strain from the QNM and surrogate models. To get the full waveforms, 
we take the strain of each mode modelled at all times and multiply by the corresponding
spherical harmonics at chosen angles $(\theta = \pi/3 , \phi = \pi/4)$. Note that the 
strains on the top panel are indistinguishable from each other by eye. However, 
the bottom panel shows the residuals between the NR waveform and our 
surrogate are $\sim 1 \times 10^{-3}$ for a QNM fitting start time of $20 M$ after peak.

\section{Conclusion}\label{sec:conclusions}

We have introduced the first ringdown surrogate model \modelName{} that
uses CCE waveforms, i.e., waveforms with memory effects, in the superrest 
frame of the remnant. Using waveforms that have been mapped to the same 
BMS frame as that of the QNM model is crucial in performing ringdown 
analyses since this is implicitly assumed for QNM solutions of the Teukolsky equation~\cite{Teukolsky:1973}.
This BMS frame fixing causes the strain to not only be shifted by a constant but also
introduces mode-mixing, as seen in Figs. 6-8
of~\cite{MaganaZertuche:2021syq}. Therefore, we fit for the QNM amplitudes 
with these CCE waveforms. Using the same fitting 
methods from~\cite{MaganaZertuche:2021syq}, we build a GPR algorithm that 
allows us to extract the remnant mass, remnant spin, and a set of 10 QNM amplitudes.
The median errors in mass and spin are $10^{-5}$, while our prediction errors for the 
complex-valued QNM amplitudes, shown in Table~\ref{tab:error_table}, are 
typically $10^{-2}$ to $10^{-3}$. These results highlight the high fidelity of our 
surrogate model for extracting high-order modes. For a visual of the ringdown 
model, Figs.~\ref{fig:strain220max}--\ref{fig:strainlmmed} show the good 
correspondence between the NR waveform and the surrogate model. 
These figures also show an example of the small residuals between NR and 
surrogate waveforms.

There are a few improvements to this ringdown surrogate model that we plan to 
implement next. Apart from overcoming the challenges with the predictions of 
retrograde and odd-$m$ QNMs (discussed in Sec.~\ref{sec:modes_modelling}), 
we plan to include mode amplitudes coming from second-order perturbation 
theory since recent works have shown their importance in precision 
modelling~\cite{Mitman:2022qdl,Cheung:2022rbm}. Additionally, we plan to model 
more general systems that may include precession. This would 
require us to fit for the spin orientation and not just the spin magnitude. 
This new model is publicly available through the python package
\texttt{surfinBH} and will continue being
updated with the improvements mentioned above. With 
these tools in hand, we will be more prepared to face gravitational-wave 
detections with a rich QNM spectrum. 

Moreover, a recent study has taken a different approach to predicting the QNMs
by finding an analytical fit for the amplitude and phase of each complex 
mode~\cite{Cheung:2023vki}. Using SXS extrapolated waveforms, they can 
recover amplitudes for the $(2,2,0,1)$ and $(2,2,1,1)$ modes with errors 
on par with our results. However, for higher harmonics, 
the fitting formula shows amplitude errors that are almost an order of magnitude 
higher than the results from Table~\ref{tab:error_table}, making it difficult to 
extract these QNMs. 

\textit{Note added.}---While preparing this article, we learned that
Pacilio et al. have built a surrogate model similar to
ours~\cite{Pacilio:2024xxx}.  We plan to validate our models against
each other in the future.

\acknowledgments
The authors would like to thank
Tousif Islam,
Costantino Pacilio,
Swetha Bhagwat,
Francesco Nobili,
and
Davide Gerosa
for helpful discussions. Some calculations were 
performed with the Wheeler cluster at the California Institute of Technology 
(Caltech), which is supported by the Sherman Fairchild Foundation and by Caltech.

The work of L.M.Z. was partially supported by the MSSGC Graduate
Research Fellowship, awarded through the NASA Cooperative Agreement
80NSSC20M0101.
L.C.S.~was supported by NSF CAREER Award PHY-2047382 and a Sloan
Foundation Research Fellowship.
K.M.~was supported by the Sherman Fairchild Foundation and NSF Grants No. PHY-2011968, PHY-2011961, PHY-2309211, PHY-2309231, and OAC-2209656 at Caltech.
S.E.F.~was supported by NSF Grant No. PHY-2110496.
V.V.~was supported by NSF Grant No. PHY-2309301. S.E.F.~and V.V.~were supported
by  
UMass Dartmouth's Marine and Undersea Technology (MUST) research program funded
by the Office of Naval Research (ONR) under grant no. N00014-23-1-2141.

\def\bibsection{\section*{References}}

\bibliography{paper_refs}

\end{document}